\documentclass[aps,amsmath,amssymb,amsfonts,prd,nofootinbib,reprint,longbibliography,superscriptaddress]{revtex4-2}

\usepackage{graphicx}
\usepackage{hyperref}

\newcommand{\dneff}{\Delta N_{\textnormal{eff}}}
\newcommand{\tn}[1]{\textnormal{#1}}

\begin{document}

\title{Dark radiation and the Hagedorn phase}

\author{Andrew R.~Frey}
\email{a.frey@uwinnipeg.ca}
\affiliation{Department of Physics and Winnipeg Institute for Theoretical 
Physics, University of Winnipeg, 515 Portage Avenue, Winnipeg,
Manitoba R3B 2E9, Canada}
\author{Ratul Mahanta}
\email{ratulmahanta@hri.res.in}
\affiliation{Harish-Chandra Research Institute, A CI of Homi Bhabha National Institute, Allahabad 211019, India}
\author{Anshuman Maharana}
\email{anshumanmaharana@hri.res.in}
\affiliation{Harish-Chandra Research Institute, A CI of Homi Bhabha National Institute, Allahabad 211019, India}

\begin{abstract}
We point out that if the sector associated with the Standard Model
degrees of freedom entered an open string Hagedorn phase in the
early universe while the dark radiation sector was not part of this
plasma, then this can lead to low values of the observable $\Delta N_{\tn{eff}}$ 
(effective number of additional neutrinolike species). For
explicit analysis, we focus on warped string compactifications with
the Standard Model degrees of freedom at the bottom of a warped
throat. If the Hubble scale during inflation is above the warped
string scale associated with the throat, then the Standard Model
sector will enter the Hagedorn phase.  In this scenario, bulk
axions are no longer dangerous from the point of view of dark
radiation.  While this article focuses on warped compactifications,
the basic idea can be relevant to any scenario where the early
universe entered a Hagedorn phase. 
\end{abstract}

\maketitle

\tableofcontents

\section{Introduction}

The hot big bang model is highly successful in addressing 
most cosmological observations. Yet, some interesting puzzles remain.
One of these is the low value of $\dneff$ -- the energy density of
new light degrees of freedom at the time of neutrino decoupling measured
as an effective number of additional neutrinolike species\footnote{In the 
nomenclature of \cite{Zyla:2020zbs}.}-- i.e., 
the absence of dark radiation \cite{planck}. Theoretically,
no reasons for a low value of $\dneff$ have been found; phenomenologically
interesting models with new light degrees of freedom including axions and 
dark photons can generate non-negligible $\dneff$. Furthermore, 
the anthropic arguments for a small value are not very strong (for
a recent discussion see \cite{Takahashi:2019ypv}). The purpose of this paper is
to point out that if the visible sector of our universe entered 
an open string Hagedorn phase \cite{Lee:1997iz,Abel:1999rq,Barbon:2004dd} after 
reheating, while
the dark radiation was not part of the thermal soup,
this can provide a mechanism of low $\dneff$. Our analysis will be 
in the context of warped string  compactifications \cite{Giddings:2001yu}, 
specifically in the setting 
of \cite{Frey:2005jk}, although the basic features of the arguments could 
well be relevant for cosmologies where
the early universe  went through a Hagedorn phase. 

Apart from containing the basic ingredients of the Standard Model (SM)
along with gravity, string compactifications typically predict various
additional species \cite{hreview}. Many of these particles can be
light. These light degrees of freedom often reside in hidden gauge
sectors which are required for consistency of the compactification
(see  e.g \cite{Giedt:2000bi, Cvetic:2004ui, Taylor:2015ppa,
Acharya:2016fge}). Generically, one also expects a multitude of
axions with a wide range of masses and decay
constants \cite{Arvanitaki:2009fg} (see  e.g 
\cite{Demirtas:2018akl, j,l1,ax} for explicit 
statistical analysis). Again, a large number of these
can be  light.  As we describe below, these light degrees of freedom
can lead to large value of $\dneff$.

$\dneff$ is  sensitive to all (relativistic) species 
that contribute to the energy density at the time of
neutrino decoupling. It is independent of  how such matter couples to
the Standard Model.\footnote{This makes  $N_{\textnormal{eff}}$ a  powerful probe of
additional hidden species.}  The ratio of the energy density in dark 
radiation to $\rho_{\tn{vis}}$, the energy density in the SM,
at the time of neutrino decoupling gives $\dneff$:
\begin{equation}
\label{basic1}
   \dneff = { 43 \over 7} { \rho_{\rm dr} (t_\nu)  \over \rho_{\rm vis} (t_\nu)}. 
\end{equation}
In scenarios where the early universe undergoes an inflationary epoch and
the constituents of the universe are produced from the decay of the inflaton, 
one can obtain a rather simple formula 
relating $\dneff$ to the branching ratios of the inflaton decay 
process\footnote{This assumes that
the energy density in the dark radiation scales as $a^{-4}(t)$ throughout its 
entire evolution; we will also discuss generalizations to cases when this is 
not true.} \cite{Cicoli:2012aq,Higaki:2012ar}
\begin{equation}
\label{basic2}
    \dneff = {43 \over 7} {B_{\rm dr} \over B_{\rm vis} } \left(  
\frac{ g (T_{\nu}) }{ g(T_{\rm rh}) }\right)^{1/3} ,
\end{equation}
where $B_{\rm dr}$ is the branching ratio for decay of the inflaton to dark 
radiation, $B_{\rm vis}$ is the branching
ratio to the visible sector, $g(T_{\nu})$ and $g(T_{\rm rh})$ the $g$-factors 
at the time of neutrino decoupling and  the reheating epoch. A similar formula 
also holds  in the case that the universe undergoes an epoch of reheating
as a result of vacuum misalignment of moduli fields \cite{mod1},  with the  
branching ratios replaced by those associated
with the decay of the light modulus (although the discussion in the present 
article will not be relevant for this case).
Note that if the ratio of the branching ratios is even of the order of one 
over five, the prediction for $\dneff$ is order
one (if the ratio of the $g$-factors is order one); conversely, if the 
ratio of branching ratios is order one, limiting to $\dneff\lesssim 1$ 
requires $g(T_{\tn{rh}})\sim 2500$, far above the high temperature SM value
$g\sim 107$. As emphasized in 
\cite{manydr}, as the topology of the compactification 
manifold becomes rich, one can expect the
number of light degrees of freedom to become large, so correspondingly
$B_{\rm dr} \sim B_{\rm vis}$ (or greater), 
which results in a large value of $\dneff$.
On the other hand, the result of observations is very different. The latest 
results from the Planck collaboration \cite{planck} give $\dneff < 0.3$,
and big bang nucleosynthesis studies give similar constraints
\cite{Cyburt:2015mya}. 
This tension has been termed as the 
``dark radiation problem in string theory.''  
Various explicit studies have confirmed this tension and proposed 
possible resolutions 
\cite{Cicoli:2012aq, Higaki:2012ar, drst, manydr, Acharya}, 
although it is fair to say that there is still no satisfactory solution. 
Furthermore, CMB stage 4 experiments \cite{Abazajian:2016yjj} should be able
to probe $\dneff \approx 0.03$. Thus developing
an understanding of scenarios with low values is  of importance from the point 
for view of future observations.

As mentioned earlier, the goal of this paper is to revisit the problem in 
a setting where the SM sector goes through a Hagedorn phase in the early 
universe. We will
do this in the context of warped compactifications making use of the setting
in \cite{Frey:2005jk}. Here, the SM is supported on branes localized at the tip
of a warped throat (or other highly warped region), and 
the warp factor provides a large hierarchy 
through the Randall-Sundrum mechanism \cite{Randall:1999ee}.
This warps the 4D effective string length at the tip of a 
warped throat to the SM scale $\ell_\tn{SM}\sim 1/M_\tn{SM}$ as opposed to the 10D
string length $\sqrt{\alpha'}$ (note, though, that $M_\tn{SM}$ may not
be the electroweak scale if warping does not provide the full SM hierarchy). 
However, during inflation, the large Hubble
scale $H\gg M_\tn{SM}$ means that the 4D effective theory would break down due 
to unsuppressed string corrections in the warped SM throat \textit{if the
SM throat remained in its vacuum state during inflation.} Instead,
\cite{Frey:2005jk} argued that the SM throat modulus is far from its true 
vacuum value during inflation due to its coupling to the inflationary sector,
leading to $\ell_\tn{SM}\sim 1/H$ during inflation.\footnote{See also
\cite{Chen:2005ad} for related discussions.} During reheating, the modulus
relaxes to its vacuum state, and the warped string scale relaxes to the
SM scale with $\ell_\tn{SM}\sim 1/H(t)$ and 4D effective field theory is just
valid throughout.

However, the total energy density available to the SM sector during reheating
is $\rho_{\tn{vis}}\sim B_{\tn{vis}} M_P^2 H^2$ (where $M_P$ is the reduced
Planck mass); a naive estimate in 
field theory assuming
rapid thermalization then gives a reheat temperature of order 
$\sqrt{HM_P}\gg H\sim 1/\ell_\tn{SM}$, so the reheating process must actually 
involve strings in the SM throat in a fundamental way. Assuming that the
SM and other sectors exchange energy slowly compared to the SM sector
thermalization time,
\cite{Frey:2005jk} gave an
argument based on entropy considerations, which we review below, 
that the SM throat reheats to a
gas of long open strings extending along the SM branes and penetrating about
a 10D string length into the warped transverse dimensions. This Hagedorn
gas has a very large effective number of degrees of freedom 
$g(T_{\textnormal{rh}})$
and subsequently decays at equilibrium primarily by emitting SM radiation 
from the string endpoints \cite{Frey:2005jk}.

We will find that this high value of
$g(T_{{\rm rh}})$ can lead to lower values of $\dneff$. The key point is
that equation (\ref{basic2}) implies $\dneff$ decreases with an increase 
in $g(T_{{\rm rh}})$.
Our analysis will show that this can significantly suppress the 
contributions to $\dneff$ from bulk light
degrees of freedom (light degrees of freedom whose  wave function is supported
in the bulk of the compactification) and also from any sector where 
the effective
number of degrees of freedom does not undergo a significant change during the
history of the universe.   Highly warped throats  (other than that containing 
the SM) with light degrees of freedom can be potentially dangerous. 
However, the problem can
be ameliorated if the number of D-branes in such throats is small or 
if the wave function of the inflation has small support in such 
throats. Before closing 
our introductory remarks, we note that highly  warped regions are 
expected to be generic in string compactifications  \cite{Warpgen}. Thus,
the article also takes an important step toward developing a full 
understanding of the nature
of predictions for dark radiation in string compactifications.

\section{The Hagedorn Phase of Open Strings}\label{s:hagedorn}

We suppose that the SM is supported on a set of D-branes in a highly warped
throat. The prototypical examples consist of either D7-branes embedded in
or D3-branes at the tip of a warped conifold throat in a conformally
Calabi-Yau manifold (with some additional structure such as an orbifold
to generate the SM gauge group), but details are unimportant for our general
argument. After inflation, the SM sector reheats to an energy density of 
about $\rho_{\tn{vis}}\lesssim M_p^2 H^2$, which is considerably above the
effective string scale. Following \cite{Frey:2005jk}, we argue that this high
energy density leads to a Hagedorn gas phase of long open strings at the
end of reheating.\footnote{At high enough energy density, there could be
a gas of black holes (as opposed to black branes) 
and strings, which has a lower temperature
and larger number of effective degrees of freedom \cite{Barbon:2004dd}. 
Our estimate is therefore conservative, and the highest temperature
phase is still a Hagedorn gas.} 
Our specific formulae below assume that the SM is supported 
on D3-branes at the tip of a warped throat, but generalizations to 
other configurations are straightforward.

Consider a pure gas of open strings stretching between D3-branes. 
The entropy of the open string gas is 
\cite{Lee:1997iz,Abel:1999rq,Barbon:2004dd}
\begin{equation}\label{entropy}
  S_o(E)=\beta_H E+\sqrt{\frac{8N^2_{D}V_\parallel E}{m\mu^2 V_\perp}}\ ,
\end{equation}
where $\mu=1/2\pi\alpha'$ is the string tension, $m$ is an order one constant
given by the probabilities for splitting and joining of long strings, 
$\beta_H=2\pi\sqrt{2\alpha'}$ is the inverse Hagedorn 
temperature,\footnote{Note that the ratio $\beta_H/\sqrt{\alpha'}$ differs
for different string theories; this is the value for type II strings
\cite{Sundborg:1984uk,Bowick:1985az}.}
$N_D$ is the number of D-branes, and $V_\parallel,\ V_\perp$ are 
respectively the volumes along and transverse to the D-branes.
For comparison, the entropy of a black 3-brane (assuming for simplicity a 
brane charge $\sim N_D$) is $S_{bb} =A\sqrt{N_D}V^{1/4}E^{3/4}$, where $A$ is
a constant of order unity. Therefore, at energy densities 
$\rho_{\tn{vis}}\gtrsim N_D^2\mu^2$, the Hagedorn phase dominates the 
microcanonical ensemble. It is also important to note that the energy density
of the open strings is much greater than that in the closed strings in 
the Hagedorn phase \cite{Lee:1997iz}.

The above comparison of entropies is of course appropriate for the 
microcanonical ensemble. With the SM localized in a warped throat region,
energy exchange between the SM and other sectors is slow due to the strong
warping, so energy deposited into the SM sector from the inflaton is 
approximately conserved, meaning the SM thermalizes within the microcanonical
ensemble. That said, the open string Hagedorn gas is also favored in the
canonical ensemble at temperatures near but below the Hagedorn temperature.

Now we can compute the inverse temperature
\begin{equation}\label{temperature}
\beta\left(=\frac{1}{T}\right)\equiv\frac{\partial S_o}{\partial E}
  \implies \beta=\beta_H+\sqrt{\frac{2N^2_{D}}{m\mu^2 V_\perp\rho}}\ ,
\end{equation}
where $\rho=E/V_\|$ is the energy density along the D-branes.
The volume transverse to the branes is not the full compactification volume;
rather, the warp factor acts as a potential preventing the strings from 
climbing the warped throat. Approximating this potential as a worldsheet
mass term for transverse oscillations, \cite{Frey:2005jk} estimated
\begin{equation}
V_\perp=v(4\pi^2\alpha^\prime)^3 . 
\end{equation}
In principle, the order 1 constant $v$ is calculable with a 
rigorous derivation of
string thermodynamics in warping, but it likely varies somewhat with the
background, so it parametrizes our ignorance of the precise compactification.
We  note that, while we have derived (\ref{entropy}) and (\ref{temperature})
in 10D units, the open strings are all localized at the tip of the
warped throat, so conversion to 4D units is simple scaling by the warp factor
at the tip. In the following, we therefore replace $\alpha'\to \ell_\tn{SM}^2$,
where $\ell_\tn{SM}$ is the warped string scale in the SM sector, and use
4D units throughout.

Therefore, we have
\begin{equation}
\beta = \beta_H+\sqrt{\frac{N^2_{D}}{8\pi^4mv\ell_\tn{SM}^2 \rho}}
\implies\ \rho= \frac{N^2_{D}}{8\pi^4mv\ell_\tn{SM}^2 (\beta-\beta_H)^2}\ .
\end{equation}
Defining $g_E(T)$ as usual by $\rho\equiv \pi^2g_E(T)T^4/30$, we find
\begin{equation}
g_E(T)=\frac{15N^2_{D}}{4\pi^6mv\ell_\tn{SM}^2}\frac{\beta^4}{ (\beta-\beta_H)^2}
=\ \frac{15N^2_{D}}{4\pi^6mv}\frac{1/(T^2\ell_\tn{SM}^2)}{ (1-T/T_H)^2}\ .
\end{equation}
Near the Hagedorn temperature,
\begin{equation}\label{gEhag}
g_E(T)\approx \frac{15N^2_{D}}{4\pi^6mv}
\frac{1/(T_H^2\ell_\tn{SM}^2)}{ (1-T/T_H)^2}=
\frac{30N^2_{D}}{\pi^4mv}\frac{1}{ (1-T/T_H)^2}\ .
\end{equation}
The effective number of degrees of freedom can therefore be very large near
the Hagedorn temperature. 

Of course, we may also measure degrees of freedom via the entropy.
At high energy densities, the entropy density is
\begin{equation}
  s_o\equiv\frac{S_o}{V_\parallel}\approx \beta_H\rho=
\frac{\pi^2}{30}g_E(T)\beta_HT^4\ .
\end{equation}
We define $g_s(T)$ by
\begin{equation}
  s_o\equiv\frac{2\pi^2}{45}g_s(T)T^3\ ,
\end{equation}
so $g_s(T)=3g_E(T)/4$ near the Hagedorn temperature.

We can also estimate reasonable values of the reheating temperature.
Parametrize the energy density of inflation as $\rho_{\tn{inf}}=a M_G^4$
in terms of the unification scale $M_G\sim 2\times 10^{16}$ GeV, where
$a$ is an order 1 (or smaller for lower-scale inflation) numerical constant,
since inflation is often taken to occur at this scale, which is also a 
reasonable proxy for the compactification and string scales and therefore
likely a maximum value for the scale of inflation.
Following the arguments of \cite{Frey:2005jk} that the modulus of the SM 
throat relaxes in such a way that effective field theory is just valid
through reheating, the Hubble parameter is 
$H=fT_H$ for $f\lesssim 1$. Since the energy density in the SM sector 
at reheating is $\rho_\tn{SM}=3B_\text{vis}M_P^2H^2\approx\pi^2g_E T_H^4/30$, 
where $B_\text{vis}\lesssim 1$, we find
\begin{equation}
\left(1-\frac{T_{\tn{rh}}}{T_H}\right)^2 \approx 
\frac{1}{3\pi^2}\frac{N_D^2}{B_\text{vis}mvf^4}\frac{H^2}{M_P^2}\ .\end{equation}
But $\rho_\tn{SM}\lesssim\rho_{\tn{inf}}$, so $H^2\lesssim aM_G^4/3M_P^2$, meaning
\begin{eqnarray}
\left(1-\frac{T_{\tn{rh}}}{T_H}\right)^2&\lesssim &
\frac{1}{9\pi^2}\frac{aN_D^2}{B_\text{vis}mvf^4}
\frac{M_G^4}{M_P^4}\ \Rightarrow\nonumber\\ 
1-\frac{T_{\tn{rh}}}{T_H} &\lesssim&
(7\times 10^{-6}) \frac{N_D}{f^2}\sqrt{\frac{a}{B_\text{vis}mv}}\ .
\label{reheattemp}
\end{eqnarray}
With $N_D=10$ for a typical D-brane embedding of the SM sector into our
string compactification, $f=1/10$, 
and $a=B_\text{vis}=m=v=1$, $T_{\tn{rh}}/T_H\gtrsim 0.993$. For a lower
scale of inflation ($a\ll 1$), this ratio is even larger.

\section{Low $\Delta N_{\textnormal{eff}}$ from a Hagedorn Phase}

\subsection{The simplest setting}

Assuming that entropy is conserved for the visible sector\footnote{We consider
entropy conservation separately for visible and dark radiation sectors.} 
between the time of reheating and the decoupling of neutrinos, we have
\begin{equation}
\label{econ}
  \left(\frac{T_\nu}{T_\text{rh}}\right)^3=
\frac{a^3(t_\text{rh})g_s(T_\text{rh})}{a^3(t_\nu)g_s(T_\nu)}\ ,
\end{equation}
independent of the equation of state ($g_s$ as well as $g_E$ below refer to
the visible sector). Therefore, for the visible sector
\begin{equation}
\label{viseqq}
  \rho_{\text{vis}}(t_\nu)=\rho_{\text{vis}}^\text{rh}
\frac{g_E(T_\nu)T_\nu^4}{g_E(T_\text{rh})T_\text{rh}^4}
=\rho_{\text{vis}}^\text{rh}\frac{g_E(T_\nu)}{g_E(T_\text{rh})}
\frac{a^4(t_\text{rh})g_s^{\frac{4}{3}}(T_\text{rh})}{a^4(t_\nu)g_s^{\frac{4}{3}}(T_\nu)}
\ .
\end{equation}
In the simplest models,  dark radiation is highly noninteracting and 
the associated effective
number of degrees of freedom do not change throughout the history of the 
universe.\footnote{This is certainly valid for very weakly interacting 
bulk axions.} The energy density falls as $a^{-4}(t)$ throughout 
the history of the universe, or 
\begin{equation}
\label{simpledr}
\rho_\text{dr}(t_\nu)=\rho_\text{dr}^\text{rh}\frac{a^4(t_\text{rh})}{a^4(t_\nu)}\ .
\end{equation}
Along the lines of \cite{Cicoli:2012aq, Acharya}, 
$\Delta N_{\textnormal{eff}}$ from \eqref{basic1} becomes
\begin{equation}
\dneff=\frac{43}{7}\frac{\rho_\text{dr}^\text{rh}}{\rho_{\text{vis}}^\text{rh}}
\frac{g_E(T_\text{rh})}{g_E(T_\nu)}
\frac{g_s^{\frac{4}{3}}(T_\nu)}{g_s^{\frac{4}{3}}(T_\text{rh})} 
=\frac{43}{7}\frac{B_\text{dr}}{B_\text{vis}}g_E^{\frac{1}{3}}(T_\nu)
\frac{g_E(T_\text{rh})}{g_s^{\frac{4}{3}}(T_\text{rh})}\ ,
\label{Nint}
\end{equation}
where we use 
$B_\text{dr}/B_\text{vis}=\rho_\text{dr}^\text{rh}/\rho_{\text{vis}}^\text{rh}$ 
(by definition), and $g_E=g_s$ at the time of neutrino decoupling. 
If $g_E=g_s$ also at reheating, we find \eqref{basic2}.
On the other hand, with $T\to T_H$ in a Hagedorn phase,
$g_s\approx\frac{3}{4}g_E$, so we can write
\begin{equation}
\Delta N_\text{eff}\approx\frac{43}{7}\frac{B_\text{dr}}{B_\text{vis}}
\left(\frac 43\right)^{4/3}
\left(\frac{g_E(T_\nu)}{g_E(T_\text{rh})}\right)^{1/3}\ .
\label{Nfin}
\end{equation}
With $T_{\tn{rh}}$ as in equation (\ref{reheattemp}), the parameters allow
for significant suppression through the ratio of $g$-factors; even 
large branching ratios $B_\text{dr}/B_\text{vis}$ can be accommodated with
intermediate or low-scale inflation, so the tension with present observations
can be addressed. Specifically, equation (\ref{Nfin}) is consistent with 
a given constraint $\dneff\lesssim \overline{\Delta N}$ for 
\begin{equation}\label{constraintTrh}
1-\frac{T_{\tn{rh}}}{T_H}\lesssim 6\times 10^{-3} N_D
\sqrt{\frac{\overline{\Delta N}^3(B_{\tn{vis}}/B_{\tn{dr}})^3}{mv}} .\end{equation}
This is similar to the estimated reheating temperature (\ref{reheattemp})
for $\overline{\Delta N}\sim 0.3$ and a ratio of branching ratios near unity
even for high-scale inflation.

\subsection{Beyond the simplest setting}

\subsubsection{Entropy production}

A key input for the above calculation was entropy conservation 
in the visible sector from the 
time that the universe was at temperature $T_{\text{rh}}$ to $T_{\nu}$. 
Another possibility is that there \textit{are} entropy generating processes 
in the visible sector
in this period (particularly early). If entropy is produced then 
equation (\ref{econ}) becomes
\begin{equation}
    a^{3} (t_\nu) g_s (T_\nu) T_{\nu}^3 = 
( 1 + \eta) a^{3}(t_{\rm rh})g_s (T_{\rm rh}) T_{\rm rh}^3\ ,
      \end{equation}
 where $\eta > 0$ i.e.,
\begin{equation}
\label{enew}
\left(\frac{T_\nu}{T_\text{rh}}\right)^3=
\frac{(1 +\eta) a^3(t_\text{rh})g_s(T_\text{rh})}{a^3(t_\nu)g_s(T_\nu)}\ .
\end{equation} 
Note that the effect of entropy production is thus captured by defining an 
effective $g^{\text{eff}}_s(T_\text{rh}) \equiv ( 1 + \eta) g_s(T_\text{rh})$. 
Now making use of this in (\ref{Nint}), in the case there is entropy 
production, we obtain the analog of (\ref{Nfin}) to be
\begin{equation}
\Delta N_\text{eff}\approx\frac{43}{7}\frac{B_\text{dr}}{B_\text{vis}}
\left(\frac{4/3}{1 + \eta}\right)^{4/3}
\left(\frac{g_E(T_\nu)}{g_E(T_\text{rh})}\right)^{1/3}\ .
\label{Nfin2}
\end{equation} 
Since $\eta > 0$, the effect of entropy production is to decrease $\dneff$, 
i.e. it helps in the issue we want to address.

We discuss entropy production in the dark radiation sector below.

\subsubsection{Nontrivial dark radiation sectors}

Let us consider the case that the dark radiation sector has nontrivial 
dynamics. As a result, equation
(\ref{simpledr}) need not be valid. As we have done for the SM sector, 
we can write
\begin{equation}
\label{drevo}
\rho_{\text{dr}}(t_\nu)=\rho_{\text{dr}}^\text{rh}
\frac{g_{dr-E}(t_\nu)T_\nu^4}{g_{dr-E}(t_\text{rh})T_\text{rh}^4}=
\rho_{\text{dr}}^\text{rh}\frac{g_{dr-E}(t_\nu)}{g_{dr-E}(t_\text{rh})}
\frac{a^4(t_\text{rh})g_{dr-s}^{\frac{4}{3}}(t_\text{rh})}{a^4(t_\nu)
g_{dr-s}^{\frac{4}{3}}(t_\nu)} \ ,
\end{equation}
where the $g_{\rm dr}$ factors are for the dark radiation sector. We have 
adopted notation
where the arguments of the dark sector $g$ factors are the times at which 
they are
to be evaluated (since dark sector can in principle be in a different 
temperature from the SM). Combining with (\ref{viseqq}) we get
\begin{equation}\label{generalDNeff}
\dneff =\! \frac{43}{7} \frac{B_{\rm dr}}{B_{\rm vis}}\!
\left(\! \frac{g_E(T_\text{rh})}{g_E(T_\nu)}
\frac{g_s^{\frac{4}{3}}(T_\nu)}{g_s^{\frac{4}{3}}(T_\text{rh})}\! \right) \!\!\!
\left(\! \frac{g_{dr-E}(t_\nu)}{g_{dr-E}(t_{\rm rh})}
\frac{g_{dr-s}^{\frac{4}{3}}(t_{\rm rh})}{g_{dr-s}^{\frac{4}{3}}(t_{\nu})}\! \right)\! .  
\end{equation}
If the dark radiation sector never enters a Hagedorn phase then the factor 
in the last brackets will be order one. 
But, if the dark sector is localized in a warped throat and enters an 
open string Hagedorn phase, then one can get a large
contribution from this factor. 

For the case that both the visible sector and the dark sector are in 
``usual thermal baths" at $t = t_{\nu}$ and in open string
Hagedorn phases at $t = t_{\tn{rh}}$, we obtain
\begin{eqnarray}
\dneff &=& \frac{43}{7} \frac{B_{\tn{dr}}}{B_{\tn{vis}}}
\left(\frac{g_{dr-E}(t_{\tn{rh}})}{g_E(T_\text{rh})}\right)^{1/3}  
\left(\frac{g_E(T_\nu)}{g_{dr-E}(t_\nu)}\right)^{1/3}\nonumber\\
&=&\frac{43}{7} \frac{B_{\tn{dr}}}{B_{\tn{vis}}}
\left(\frac{g_E(T_\nu)}{g_E(T_\text{rh})}\right)^{1/3}  
\left(\frac{g_{dr-E}(t_{\rm rh})}{g_{dr-E}(t_\nu)}\right)^{1/3}\ .
\label{bothhag}\end{eqnarray}
A large contribution can arise from the factor 
$$
\left(g_{dr-E}(t_{\tn{rh}}) \right)^{1/3}  \sim  
\frac{N^{2/3}_{\tn{dr-D}}}{(1 - T_{\tn{dr-rh}}/T_{\tn{dr-}H})^{2/3}} 
\left( \frac{T_{\tn{dr-rh}}}{T_{\tn{rh}}} \right)^{4/3}
$$
where $T_{\tn{dr-}H}$ is the Hagedorn temperature  in the dark sector,
$T_{\tn{dr-rh}}$ the temperature of the 
dark sector at $t_{\tn{rh}}$ and $N_{\tn{dr-D}}$ the number of D-branes
in the dark sector. This is the case when the dark radiation is, for example,
an unconfined gauge theory on D3-branes in a separate warped throat,
although the contribution is suppressed if the number of D-branes in
the dark throat is much less than that in the SM throat. Another possibility
for the suppression of the contribution is that the inflaton branching
ratios to  throats which contain dark radiation candidates are low, which 
can happen if the inflaton wave function
in the extra dimensions has limited support in the dark radiation 
throat.\footnote{This is
possible as throats occupy localized regions of the compact dimensions.}

Specifically for axions, even if these are localized in a strongly 
warped region, we do not
expect them to gain this sort of Hagedorn enhancement. If there are no
D-branes present, a similar argument to that of section \ref{s:hagedorn}
shows that the black 3-brane dominates the closed string Hagedorn gas in
the microcanonical ensemble in the thermodynamic limit (of fixed energy
density and infinite parallel volume). Since the black 3-brane has the same
equation of state as radiation, we do not expect a parametric enhancement
of degrees of freedom.

We can also consider entropy generation in the dark radiation sector. 
If the fractional entropy increase is $\eta$ (that is, equation (\ref{enew})
applies for the dark radiation sector), that modifies equation 
(\ref{generalDNeff}) by inserting a factor of $(1+\eta)^{4/3}$ in the 
numerator, which increases $\dneff$.

\section{Conclusion and Discussions}

We have presented a proof of concept that an early open string Hagedorn 
phase can alleviate
the dark radiation problem of string theory.  We focused on the setting of 
\cite{Frey:2005jk}, where the SM is 
localized in a warped throat, with the warped string scale below the Hubble 
scale during inflation. At the end of
inflation, the SM  sector enters a Hagedorn phase. The effective number
of degrees of freedom in this phase is very high, diverging as $(T - T_H)^{-2}$
as $T \to T_H$. We found that for dark radiation which is not part of the
thermal plasma (such as bulk axions) this large effective number of degrees
of freedom leads to suppression in $\dneff$. We also discussed
the effects of entropy production and dark radiation sectors in other throats.
In summary, the idea provides an attractive way to address the dark 
radiation problem
in string compactifications. Furthermore, since warping is a generic feature 
of string
compactifications, the work provides an important element for acquiring 
a complete
understanding of the nature of predictions for dark radiation in string theory.
   
Our basic idea extends to many other models of early universe cosmology in
string theory. For example, it seems likely that an open string Hagedorn phase
appears in any model where the SM (or sufficiently strong intermediate) 
hierarchy is the result of warping or other inhomogeneity of a string 
compactification with the SM supported on D-branes. However, the Hagedorn
phase of strings also appears intrinsically in the string gas model of 
cosmology \cite{Brandenberger:1988aj,Nayeri:2005ck} (or generalizations
including black holes at the correspondence point with strings
\cite{Quintin:2018loc}), which is of current
interest in light of conjectured constraints on inflation in the swampland
program \cite{Obied:2018sgi,Bedroya:2019snp}. While the original string
gas cosmology exclusively discusses closed string degrees of freedom, there
is a natural extension involving branes \cite{Alexander:2000xv}; since a 
dilute gas of D-branes converts closed strings to open strings efficiently,
it is reasonable to expect that the energy density in a brane-supported SM
would dominate over closed string dark radiation in a variation of the
mechanism described here. Similar comments might apply to the models of
\cite{Agrawal:2020xek} given their conceptual relationships to string gas
cosmology. One difference with the key ideas presented here is important:
we have considered a scenario in which the SM and dark radiation sectors
are sequestered, but that is not the case in the string gas cosmologies
studied so far (and may not be the case in more general warped 
compactifications). As a result, the dynamics of the transition between 
the open string Hagedorn and radiation phases, including coupling to the
closed string sector, could play a more important role in a more general
class of models.

Of course, there are several other interesting directions for future work. 
Let us list a few. First, one can
try to embed the scenario in a concrete model of moduli stabilization. This will
allow for explicit computations of various parameters and will help to arrive at
explicit numerical values for predictions for $\dneff$. Second, the present
work relied on the understanding of the open string Hagedorn phase in 
warped throats as developed
in \cite{Frey:2005jk}. There are many fronts in which this can be improved 
as described in detail in \cite{Frey:2005jk}.
Finally, one can explore whether the Hagedorn phase has any other characteristic
signatures for observations and how these correlate with $\dneff$.

\begin{acknowledgments}
RM is supported in part by the INFOSYS scholarship for senior students (HRI). 
AM is supported in part by the SERB, DST, Government of India by the 
grant MTR/2019/000267. ARF is supported by the Natural Sciences and
Engineering Research Council of Canada Discovery Grant program, 
grant number 2020-00054.
\end{acknowledgments}

\end{document}